\documentclass[preprintnumbers,amsmath,amssymb,nofootinbib
,superscriptaddress]{revtex4}
\usepackage{hyperref}
\usepackage{graphicx}
\usepackage{color}
\usepackage{xcolor}
\usepackage{comment}
\usepackage{lmodern}

\newcommand{\be}{\begin{equation}}  
\newcommand{\ee}{\end{equation}}  
\newcommand{\bear}{\begin{eqnarray}}   
\newcommand{\eear}{\end{eqnarray}}  
\newcommand{\ba}{\begin{array}}  
\newcommand{\ea}{\end{array}}

\usepackage{diagbox}

\definecolor{rossoCP3}{cmyk}{0,.88,.77,.40}
\definecolor{blueRef}{rgb}{0.2,0.2,0.6}
\definecolor{blue}{rgb}{0,0.396,0.741}
\hypersetup{
	colorlinks, 
	bookmarksopen, 
	bookmarksnumbered,
	citecolor=blueRef, 		
	linkcolor=rossoCP3,	
	urlcolor=rossoCP3,			
}

\newskip\humongous \humongous=0pt plus 1000pt minus 1000pt

\newif\ifdtup

\def\oldreffmt#1{\rlap{[#1]} \hbox to 2\parindent{}}

\def\figfmt#1{\rlap{Figure {#1}} \hbox to 1in{}}  
  
\def\beq{\begin{equation}}  
\def\eeq{\end{equation}}  
\def\bea{\begin{eqnarray}}  
\def\eea{\end{eqnarray}}

\def\bq{\begin{quote}}  
\def\eq{\end{quote}}

\relax

\newdimen\tdim  
\tdim=\unitlength  
\def\bar{\overline}

\begin{document}

\title{William A. Bardeen $-$ A Brief Biography {\small \cite{source} } }

\author{Christopher T. Hill}
\email{hill.delafield.physics@gmail.com}
\affiliation{Emeritus, Fermi National Accelerator Laboratory,
P. O. Box 500, Batavia, IL 60510, USA}
\affiliation{Honorary Fellow, Department of Physics, University of Wisconsin-Madison, Madison, WI, 53706
}

\begin{abstract}
\vspace{0.1 in}
William Allan  Bardeen  (September 15, 1941 $-$ November 18, 2025) was an American theoretical physicist who worked at the Fermi National Accelerator Laboratory.  He is renowned for his foundational work on the chiral anomaly, the Adler-Bardeen theorem,
the non-Abelian anomaly and gravitational anomalies. He was instrumental in the development of quantum chromodynamics and its applications, such as semileptonic decays and the $\Lambda_{\bar{MS}}$ scheme frequently used in perturbative analysis of high energy processes involving strong interactions.  Bardeen also played a major role in developing a theory of dynamical breaking of electroweak symmetry via top quark condensates, leading to one of the first composite Brout-Englert-Higgs boson models.  His work on the chiral symmetry dynamics of heavy-light quark bound states correctly predicted abnormally long-lived resonances which are chiral symmetry partners of the ground state.
\end{abstract}

\maketitle
 
\date{\today}

\email{hill.delafield.physics@gmail.gov}


\vspace{0.1 in }

Bill Bardeen was born in 1941 in Washington, Pennsylvania.  At the time, the family was living in Fairfax Village, Virginia, where his father, John, worked at the Naval Ordinance Labs. When the war was over, they moved to Summit, New Jersey, where his father joined Bell Labs.  John Bardeen, with Walter Brattain and William Shockley,  invented the transistor in 1947 at Bell Labs (for which they received the 1956 Nobel Prize in Physics).   
In 1951 the family relocated to Champaign-Urbana,  Illinois, where John Bardeen  took  an academic position at the University of Illinois.  There,  Bill attended the University High School which had a special five-year program that combined seventh and eighth grades into one year, followed by the usual freshman, sophomore, junior, senior sequence.  Bill recalled, in particular,  his high school chemistry teacher, Mr. Westmeyer,  was  a great inspiration to his students.     

Bill  had an early interest in electronics and hands-on science.  With his friend,  John Muntyan, he would bicycle into town where a war surplus store provided cheap components to build electronic devices.  Bill was handy at fixing things, such as rebuilding the lawn mower so it worked better, or fixing the dishwasher, etc.   As a high school senior he participated in the Westinghouse Science Talent Search with a project of making a furnace in the basement where he could study the mobility of color centers in crystals.   His father had two positions, one in the Physics Department where he worked on the theory of superconductivity (which led to the Bardeen-Cooper-Schrieffer theory and John Bardeen's second Nobel Prize) and another position in the Electrical Engineering Department where he had a semiconductor lab.  Bill enjoyed ``hanging out'' in the  lab, and scrounging parts such as some early transistors.  The lab was run by Nick Holonyak, his father's first graduate student and the inventor of Light Emitting Diodes (LEDs), who was very happy to encourage Bill's interest in electronics.  Bill also had a summer job, both in high school and college, in a biophysics lab where he earned about \$0.75  an hour.  There he helped study the wiring of cat brains and frog muscle response to ultrasound.

Bill's older brother, Jim, went into theoretical astrophysics, studying black holes at Harvard and Caltech, where he was a student of William Fowler and Richard Feynman.   While Bill was a sports enthusiast and enjoyed building things,  Jim was a more quiet and cerebral person, like his father John, though he was very interested in science fiction.  His younger sister, Betsy, was talented in music and  mathematics and followed a career in developing large scale database software.  She married Tom Greytak, who studied Bose-Einstein Condensation as a professor of physics at MIT.    Surprisingly, however, when the family got together, physics was not the main topic of discussion.

Bill met his to-be wife, Marge,  in high school chemistry class.  They both went to Cornell University, where they married when they were still undergraduates.  Their son, Chuck, was was born in Ithaca Dec 16, 1961.   Bill graduated in 1962 and moved on to graduate school at the University of Minnesota.  At the time, Minnesota had several good departments but he ``flipped a coin'' and ended up in particle physics.   He received a NASA Traineeship to support his graduate studies from 1963-1966 so he no longer had time consuming teaching assistantship duties and was free to focus on research.  Steve Gasiorowicz became his thesis advisor.
Gasiorowicz  had been at Minnesota only a short time when Bill arrived, was very well-informed about the physics literature and had many professional connections around the country.  He was  busy writing a textbook on particle physics \cite{Gas}.  At Minnesota there was an active young group of both students and postdocs, and while he didn't collaborate when he was in graduate school, Bill was nonetheless inspired by the many interactions with other people.  

Their daughter, Karen, was born in Minneapolis May 6, 1965.  Expecting to finish his thesis he applied for postdoctoral positions in 1966.  One of the places he applied to was Stony Brook with the new Institute for Theoretical Physics, headed by C. N. Yang (Frank Yang).   Bill received the offer and  Gasiorowicz said, ``Take two minutes and accept.''  

Although Bill had finished the course work in graduate school he hadn't quite completed his Ph.D. thesis. He stayed a little longer at Minnesota to finish that and then the family moved to Stony Brook.  The thesis was about three quarters completed when he left for Stony Brook where he thought he would do his postdoc work in the daytime and thesis in the evening. However, it took another year and a half to finally finish the thesis and it had become an entirely different topic than he had originally begun.   In 1968, he returned to defend his work at Minnesota and was awarded his Ph.D. after he had already accepted his second postdoc at the Institute for Advanced Study in Princeton (later on, some people advised him that it might have been better not to bother to get a Ph.D. degree, like Freeman Dyson, a famous physicist who didn't have one).  Fortunately, Stony Brook was not so bureaucratic that they wanted the paperwork for all of his previously earned degrees before hiring him. (He recalled that when he applied to Stanford for a faculty job they required paperwork starting with high school transcripts!)

At Stony Brook there was also a young and enthusiastic group.  Ben Lee was a senior faculty member,   Boris Kaiser was an assistant professor.  There were four postdocs, Bill, Hwa-Tung Nieh, Wu-Ki Tung, Michael Martin Nieto and they were all put in one office, four desks in one room, which encouraged interaction.   When he went to Stony Brook in 1966, Bill was inspired by Steven Weinberg's work on current algebra and chiral Lagrangians.  He worked on applying these ideas to a wide variety of problems, some of which was subsequently coauthored with Ben Lee  and others, and there were visits by Lowell Brown which were inspiring because he was very interested in the same problems \cite{Lee1,Lee2}.  He also worked with Wu-Ki Tung on calculations in perturbative quantum field theory \cite{Tung}.  At that time, Julian Schwinger visited when he was developing his ``Source Theory''  and  gave a series of lectures. Bill  was rather skeptical about this new approach.   Schwinger was  ``a very smooth lecturer,''  but the simplest and most difficult parts of the talk were given with the same rhythm.  Bill remarked,  ``If you weren't awake, you would find out that he made a big step, and you didn't catch it.'' Yet, it was inspirational to hear from such a famous physicist.

In 1968, Bill immediately received  and accepted an offer from the Institute for Advanced Study  (IAS) in Princeton.  The IAS was a unique environment with a large group of postdocs in various fields and certainly in particle physics.  Various families lived together in the IAS Member housing just across the street from the office buildings.  It was easy to go back to work when following a good idea.  Two young long-term IAS members, Steve Adler and Roger Dashen, had recently arrived at  the Institute and lived in the housing.  This made for an interesting social life as both Steve and Roger were very enthusiastic people.  While there were some famous older faculty, Bill interacted more with the other postdocs and with Steve and Roger.

\vspace{0.25 in}

At the IAS, Bill continued to work on chiral Lagrangians and the properties of currents that were important for the weak interactions and current algebra.  Adler had already done seminal work on chiral anomalies \cite{Adler}, and inspired Bill in the general properties of chiral anomalies.  The anomaly represents a special situation in quantum theory: the vector and axial vector currents are conserved in the classical theory (tree Feynman diagrams),  but the quantum chiral anomalies imply that they don't both stay conserved (at the level of Feynman loops) because they then conflict with each other.  
In 1969, at the IAS and building upon Adler's earlier work,  Bill computed the most general form of the chiral anomalies in four dimensions for vector and axial vector currents, including the case of Yang-Mills theories \cite{anomaly1}.  This  analysis was a {\em tour de force} and, moreover in the same year,  Bill and Adler proved that the anomalies  in the  one loop calculations were not modified by higher order corrections, a celebrated result known as the ``Adler-Bardeen theorem'' \cite{anomaly2}.  Therefore, the anomaly structure was an exact result in quantum field theory.\footnote{E.g., in QED after renormalization of couplings and fields, the anomaly coefficient remains the same as in the one-loop calculation:  So $\partial j^5 =  c e^2 F \widetilde{F}$, under renoomalization becomes, $ \longrightarrow  \partial j'{}^5 =c' e'{}^2 F' \widetilde{F}'$, and  then the Adler-Bardeen
theorem states that $c = c'$. }

Bardeen's anomaly paper,  ``{\em Anomalous Ward identities in spinor field theories}" \cite{anomaly1},   is of foundational importance to modern physics.  This led to the distinction between the ``consistent anomaly" and the ``covariant anomaly."   The covariant anomaly pertains to physics, e.g., through the ``triangle diagram'' for the decay of the neutral $\pi^0$ meson into two photons.  In QED this is essentially the original calculation of Steinberger from 1949 \cite{Steinberger}, and the recognition that this is a quantum violation of chiral symmetry by Bell and Jackiw \cite{Bell}.  

The trickier calculation is the amplitude:  (axial current) $\rightarrow $ 2 photons, where one then takes the divergence of the current, which is the analysis of Adler \cite{Adler}.  This requires care with counterterms, that are needed to fully define the Feynman loop diagrams, and which Adler discussed in detail.  If the vector current is conserved, as required by gauge invariance in e.g. quantum electrodynamics, the counterterm explicitly implies that the axial current conservation (chiral symmetry) is broken by the anomaly.  However in a theory with a pure left-handed fermion there is only one current, and the anomaly is generated with the result $(1/3)\times$ that of Adler \cite{anomaly1}.   This is the ``consistent anomaly."\footnote{Note that a single left-handed electron coupled to electrodynamics becomes  an inconsistent theory due to the anomaly,
since $\partial_\mu F^{\mu\nu} = e j^\nu$ hence $\partial_\mu \partial_\nu F^{\mu\nu} = 0 = e \partial_\nu j^\nu$, so
the current must be conserved by the antisymmetry of $F^{\mu\nu}$.}

Bardeen completed his analysis, relying upon brute calculational force to get the general results. In Yang-Mills theory one must compute triangle, box and pentagon loops. The covariant anomaly for the singlet current in a Yang-Mills theory involves introduction of what is now called the ``Bardeen counterterm,''  and the resulting anomaly is a gauge invariant function of the Yang-Mills fields.  However, the consistent anomaly for a non-singlet current (which Bardeen referred to as the ``left-right symmetric anomaly'') is not  gauge invariant.  This seemed mysterious in Bardeen's calculation and it wasn't clear what it meant. Moreover,  Bardeen's paper was disputed by a referee.  If the result is not manifestly gauge invariant what then governs it?

Subsequent to Bardeen's paper it was shown that the consistent anomaly is a definition of the loop integrals, via counterterms, that ultimately satisfies ``Wess-Zumino consistency conditions'' \cite{Wess}. The paper of Wess-Zumino appeared after Bardeen's paper and it resoundingly validated Bardeen's result, so his paper was finally published.  The  Wess-Zumino consistency conditions imply that the consistent anomaly is related to topology, e.g., the structure and coefficient obtained by Bardeen's calculation turns out to be equivalent to that generated by a topological ``Chern-Simons term'' built of Yang-Mills fields in 5-dimensions on the 4-dimensional boundary.  So, where gauge invariance failed, topology governs the anomaly!\footnote{In general, anomalies occur in even dimensions, while Chern-Simons terms occur in odd dimensions, and the anomaly is the boundary of the Chern-Simons term one dimension higher (this is called ``descent cohomology'').} 

Bardeen's result anticipated the connection of anomalies to topology, as well as physics.
Mathematicians were interested in this work because physics was displaying the relationship with various topological structures in mathematics which had been discovered a few years earlier.  It is also essential to represent the anomaly which governs, e.g.,
$\pi^0$ decay and other processes, in chiral Lagrangians of
spin-0 and spin-1 mesons, which is the role played by the Wess-Zumino-Witten term \cite{Wess}\cite{Witten},
together with the relevant counterterms \cite{Rajeev}\cite{HHH}. The anomaly is also fundamental to instantons and
the solution of the $U(1)$ problem of QCD, discovered by 't Hooft \cite{thooft}.
The work on chiral anomalies is probably Bill Bardeen's most important contribution and it underlies considerable activity in the subsequent years with the resurgence of quantum field theory, the standard model and its theoretical extensions. Understanding its context in both mathematics and quantum field theory has been important over the past 50 years of theoretical particle physics.

\vspace{0.25 in}

 Normally, a postdoc at the IAS would be expected to have a two year position but, during Bill's first year at the IAS, his former Stony Brook colleague, Frank Yang, recommended him for a faculty position at Stanford University.  Learning of this, Bill formally applied to Stanford in 1969 but he didn't expect to get the job (given a history of rejection by Stanford for earlier undergraduate and graduate student positions).  However,  he did get the offer which led to a bit of a quandary:  He hadn't applied any place else because he wasn't really ready to apply for a faculty job.  He had to decide to go to Stanford, or perhaps, somewhere else, but Stanford expected him to accept immediately on the phone.  He  managed to defer for a week, but other jobs didn't materialize that quickly.  He wasn't sure what to do since he was quite happy at the IAS.

There were stories about Stanford having conflicts between the university, where he was offered the assistant professorship, and SLAC (Stanford Linear Accelerator Center), which was one of the younger national labs dedicated to particle physics research.  The clash between some of the faculty members made him a little nervous about going,  but in the end he accepted the offer.   Prof. Leonard Schiff, who had championed the offer,  died a year or so after Bill went there.  The theoretical physics group was led by Dirk Walecka who was a nuclear theorist.  Three assistant professors, not usually considered to be tenure track positions, were particle theorists.

On the other hand, SLAC had a large active group in theoretical particle physics with both junior and senior staff, where the senior staff was led by Sid Drell and James Bjorken.  While Bill worried a little bit about the tension between Stanford and SLAC,  he chose not to be part of it.  He had an office on campus, and never managed to get a desk at SLAC, hence he didn't  spend a lot of time there.  At Stanford he was teaching and found this to be a considerable effort and very time consuming, which  also didn't encourage him to spend time at SLAC.   (Some people may be amused that one would complain about teaching one course a quarter where a community college professor teaches three or four; Bill's daughter became a science teacher and taught five courses every day at Oak Park River Forest High School, rather than one course, three days a week).  In any case teaching courses was new to Bill, and it took a lot of time and he felt that he wasn't that good at it.

In terms of research, Bill continued to work on properties of currents and the nature of weak and electromagnetic interactions and the possibility that they could be described by a renormalizable quantum field theory.   He  also worked with several people at SLAC on something called the ``SLAC Bag Model,'' \cite{Bag}  (there was  also an ``MIT Bag model,'' \cite{Bag2}),  which is how one might describe confined quarks,  i.e.,  quarks that are the substructure of neutrons and protons, though free quarks cannot be produced in the lab and are always localized inside the particles we do see.  Bill commented,  ``I would say at any given time, there was one enthusiast, three sort of middle of the road, and somebody who thought it was all completely crazy. Who was enthusiastic and who was negative, varied from time to time, so it was a dynamical collaboration.''  

\vspace{0.25 in}

At this time there was growing excitement about  the theory of unified weak and electromagnetic interactions, due to Glashow, Salam  and Weinberg.  Gerard 't Hooft had shown that Yang-Mills gauge theories were renormalizable.   Many people were looking around for definitive things to calculate now that theories were indeed calculable.  Bill worked with Raymond Gastmans and Benny Lautrup on Feynman loop corrections to various electroweak processes (that ultimately had an interesting connection to the $g\!-\!2$ of the muon, measured precisely at Fermilab \cite{gastmans}).

Bill received an Alfred P. Sloan Foundation Fellowship which paid for him to spend a year at CERN in 1971-72.   There he worked with a wide variety of people, in particular with Harald Fritzsch and Murray Gell-Mann who were both visiting CERN that year.  Murray would come and go;  Harald was there most of the time.  They were both interested in the nature of quarks, in particular the evidence for dynamical quarks that carried the quantum number dubbed ``color.''  They considered a novel Yang-Mills  theory of quark interactions, in which each quark comes in one of three varieties of colors, e.g. ``red, blue and yellow.''  They wrote down the Lagrangian of the theory now known as quantum chromodynamics (QCD).  Through the chiral anomaly and Adler's $\pi^0$ decay calculation, Bill established the existence of the three quark colors from the measured rate of $\pi^0$ decay \cite{pidecay}.

At CERN Bill also continued to work on chiral anomalies and their role in determining the structure of the weak and electromagnetic interactions.  He became interested in ``K-mesic atoms,'' and consulted with an experimental group at CERN  that was making the measurements.  While they were motivated by measuring the neutron distribution, Bill showed that most of the K-mesons get captured on protons and therefore were only sensitive to the proton distribution.  Of course, the group was disappointed in this, but they combined his theoretical work with their analysis programs and showed that the predictions agreed very well with what they were observing and disagreed with the conventional analysis of capture on neutrons.

During this time, Bill spent a month at the Max Planck Institute in Munich and attended an important conference in Paris during the spring or summer on renormalization of electroweak theories, where there was a lot of debate about anomalies.  Bill was one of the first who observed that the gauge anomalies canceled in a full generation of standard model fermions allowing the renormalizability.  During a lecture by 't Hooft  it was suggested the anomaly cancellation may not be required for the theory to be renormalizable.  In the debate that ensued, Bill went to the projector screen and wrote some Feynman diagrams showing that one definitely has to cancel the anomalies, otherwise they mess up the gauge symmetry of the theory at higher order.  The cancellation of anomalies thus became a hallmark  of the standard model and its extensions.

Bill was offered a faculty position in Nijmegen, Belgium, and  received an offer to stay for a longer term at CERN.  He considered it, but felt he had displaced his family for a year and wasn't sure he wanted to move permanently to Europe. He still had an assistant professorship at Stanford but with the expectation he wouldn't be promoted.

\vspace{0.25 in}

Part of the CERN experience was living in Switzerland where he rented an apartment from Maurice Jacob, then a CERN staff member, 
in Grand-Saconnex. There they sent the kids to the local schools where Karen was in first grade and Chuck was placed in fourth grade (although he would have been in fifth grade in Palo Alto).  Karen was very quiet in the classroom until after Christmas when she felt comfortable speaking French. She first learned to read in French.  Chuck won the prize for the student who worked the hardest and would have been double-promoted to sixth grade if they had stayed on.  

They all learned to ski and discovered the trick of weekend skiing in the Juras: The Swiss students went home from half-day classes for a midday meal on Saturday that was the main meal of the day, and the Bardeens then went to the mountains, had a crepe or a hot dog, and hit the slopes, then hot chocolate and a doughnut on the way home, and acquired a taste for raclette and fondue. They learned to read Asterix and Obelix, Tintin and Lucky Luke cartoon books, in French.  

 Once, they noticed there was going to be a local basketball game in the school gym and thought it would be fun to attend for a little American culture.  Bill approached one of the team members and asked if he could be on the team.  Sure enough, he could --- and he proved to be one of the better players and a good shooter.  They went to all the home games and once or twice were invited to join the group at the home of one of the players.  At the end of the season the team played in a tournament where teams came from all over Switzerland and were of different levels and Bill's team was in Level C.   They started at eight in the morning.  If the team won, they played in the next round.  Bill phoned Marge in the middle of the day to say they kept on winning so he would have to  stay for more games.  By the time they played four games they were in the finals, and dead tired.   The team they were then playing had good players, some who were  6' 8'' height, so they didn't do very well in the final game, but the team did move up to Level B.   Some of his teammates said they had official connections, and they could get him a regular job in Switzerland!   Getting a residence permit in Switzerland isn't so easy, but they were sure they could do it if he wanted to stay.  So, he had three offers to stay in Europe, but, in the end he went back to Stanford.

 When Bill returned to Stanford, he was promoted to Associate Professor but still without tenure. The  next year, he thought he better start looking for jobs. Congress had passed the Mansfield Amendment late in 1969. The amendment barred the Defense Department from using its funds ``to carry out any research project or study unless such project or study has a direct and apparent relationship to a specific military function."  Many universities after World War II had been funded by grants from the Navy, the Army, the Air Force (Bill's father had Navy contracts).  Probably, from the Los Alamos experience, the Armed Forces felt it was important to maintain connections with scientists and understood that doing basic science was important for progress, even though it wasn't immediately of help the military.  However, with multiple  protests against the Vietnam War at universities, the Mansfield Amendment intended to make non-military funding available to the universities.  This meant that many of the university groups that had been supported under military contracts lost their funding, and there was a scramble to replace that funding with support from the National Science Foundation and Atomic Energy Commission.  This couldn't happen instantaneously so, there ensued a very tight period for the funding for university positions. When Bill was looking for jobs in 1973 and 1974, the market was very tight.

 \vspace{0.25 in}
 
In the following year, 1974, Bill received an offer to visit Fermilab which was the  new ``energy frontier'' laboratory, and it had a new theory group.  There had been an earlier temporary theory group with a temporary head and some postdocs.  In  1974, Ben Lee and Chris Quigg came from Stony Brook to start the permanent theory group at Fermilab.  Bill came for a six-month visit in summer and fall of 1974.  Because he had previously worked with Ben Lee  he thought it might be enjoyable to work with him again.  In the interim he went back to Stanford to teach another two quarters.

Stanford was considering whether to promote or not. Usually, the faculty position that Bill held was not considered eligible for promotion because they wanted to keep young people flowing through the department. If they promoted him, they would have fewer assistant professors, so they thought about it for a long time. In the meantime Bill was having ups and downs: One week he thought, Fermilab likes me and is a very attractive place to be; but the next week he thought Stanford and SLAC were a good place to be.  Finally, he received the letter from Stanford promoting him to Associate Professor with tenure, but that week he was leaning toward Fermilab.

So the Bardeens moved back to Illinois and Bill went to Fermilab.     Like the ITP at Stony Brook, Fermilab was a new group when Bill came in Fall of 1975.  Again,  there were people from around the world who came to Fermilab as postdocs which made for a lively group.  Bill collaborated with Ben Lee on problems in field theory and worked with some of the postdocs. While he was still at Stanford, Bill had received an offer for a von Humboldt Foundation Fellowship.  He couldn't use it right away because he had just arrived and thought he'd better integrate into Fermilab.  He decided to go to the Max Planck Institute in Munich for six months and use the von Humboldt in 1977.
 
Tragically in 1977 Ben Lee died in an auto accident driving out to Aspen.  This was a big shock to the theoretical physics community and, in particular, the Fermilab group.  Chris Quigg became the Head of Theoretical Physics and made a heroic effort keeping the group together and functioning.

Bill continued to pursue research in QCD, where ``gluons''  
are combined with quarks to construct the field theory that describes the strong interactions. So how could one apply the theory to realistic problems?  Here the issue for QCD is to
precisely define the theoretical inputs.  For e.g., quantum electrodynamics, the main input is the value of 
the electromagnetic coupling, $\alpha$ (together with particle masses, etc.). However,
in QCD the coupling $\alpha_{\mathrm{QCD}}(\mu/\Lambda_{\mathrm{QCD}})$ is changing (``running'' fast) with scale $\mu$.  Hence the relevant input became $\Lambda_{\mathrm{QCD}}$, and this must be done with a well-defined procedure to systematically accommodate radiative corrections. 
Bill wrote a seminal paper with Andrzej Buras, Dennis Duke and Taizo Muta, \cite{Buras1} which introduced a suitably
defined $\Lambda_{\mathrm{QCD}}=\Lambda_{\bar{\mathrm{MS}}}$. They then worked out how to systematically apply this to any higher order QCD corrections to processes measurable in the laboratory.  This also led to  a specific application to the photon structure function \cite{Buras2}.

Other works of that era include continued collaboration with Ben Lee and Robert Shrock \cite{Lee} on phase transitions,
and with Henry Tye \cite{Tye}, exploring current algebra and a low mass Higgs boson. 
Working with Bruno Zumino,  Bill extended anomalies into the general structure of gravity and string theory
 which lead to a highly cited paper that is also an excellent tutorial on the algebraic and local symmetry structure of general relativity \cite{anomaly3}.  He also did collaborative work with Sherwin Love, Chung Ngoc (Terry) Leung  \cite{Leung}, on the possible role of scale invariance in particle physics.  Bill produced many other papers
 with Andrzej Buras over many years, mainly on applying QCD to physical processes such as developing the large-$N_{\mathrm{color}}$ 
 expansion in QCD, with applications to K-meson physics \cite{Buras3}.  

 \vspace{0.25 in}
 
 Bill  recalled that when they first came to Fermilab, he and Marge were invited to dinner at Dick Carrigan's house:

 \begin{quote} ``Bob Wilson was there, and he went on at some length about how useless theorists were. That was my welcome to Fermilab.  I would say we were tolerated by Bob Wilson and encouraged by Ned Goldwasser who was the Deputy Director. In a way, the theory group started before there officially was a theory group because the theoretical community felt it was important that Fermilab, a new laboratory, have people to talk about theory. In the beginning, it was postdocs and a rotating Head visiting from different places.

 Establishing the theory group may have come from outside pressure rather than inside pressure. Various theorists have had different kinds of connections with the experiments. I would say I probably didn't have that strong a connection except socially. When we moved here, I carpooled with some of the experimentalists. We would talk what's going on in experiment and theory. But later on, I was certainly on committees of various kinds. But some of the other members of the group, in particular Chris Quigg, were much more directly involved in talking with the experimentalists.

There was a difference between being an experimental physicist at Fermilab and a theoretical physicist because as a theorist, your main job was to do inquiry-based theory. Whereas the experimentalist had a requirement to spend part of their time supporting the experiments, a research fraction. He or she had a real job making something work and then some research time to work on an experiment. Usually, the research time was in the evenings, if at all. (Theorists work in the evenings too!)

I don't think Bob Wilson's attitude particularly affected how the theory group functioned. He didn't put us on a pedestal, but he didn't get in the way.  I think he was more interested in building the laboratory. And, of course, he was often fighting Washington and ultimately resigned over the fact that he couldn't get the government to see things his way. Leon [Lederman] was much different. He liked to talk to theorists. However, he would tell theory jokes that were usually at the expense of theorists, but they were jokes. I took them that way.   Certainly, the theory group has evolved. When I came, the group had a wide variety of interests in inquiry-based research.''
\end{quote}

In 1987, Chris Quigg left Fermilab to work on the Superconducting Super Collider (SSC), and Bill became Head of the group.  Bill had to decide about promoting junior people.  He described his management philosophy:

\begin{quote} ``The question was, do we continue to try to maintain a strength in very formal areas of theoretical physics, or should we focus more on areas related to experiment?  We decided that we should focus more on inquiry-based research related to experiment.  More recently, lab research has become less-inquiry based and more project-oriented. In other words, there are subgroups in QCD phenomenology, Lattice Gauge Theory and Beyond the Standard Model Physics.  The DOE wanted to classify people that way.

The Fermilab Theory group would always review hundreds of applicants for postdocs, or even junior faculty, and  try to select the best available people.  Nowadays,  it tends to be the best available person in a subgroup,  e.g., Lattice Gauge Theory, or  QCD phenomenology.  This has  given the group a different character.  When I retired I was invited to continue to participate in hiring discussions, but I felt strongly that ``old folks'' should not interfere with young people finding their way. I had been in discussions like this for 35 years. That's long enough.

I always wanted to find and encourage the best people to come to Fermilab, to maintain a very active group of young people, and to encourage the senior staff to interact with the younger people. Sometimes, when people joined the group, they wanted to spend a day a week at home (something which now seems to be encouraged).  I felt part of the job as a senior person was to interact with young people. If you weren't here, you couldn't do that. Of course, back in those days, we didn't have the Internet and the ways people communicate today. It was a different time. If you look at the basics, we wanted the best people, we wanted some diversity, and we wanted a variety of interests and viewpoints. Once we selected the top 20 candidates, we looked for different talents. We didn't want to hire everybody to be cookie-cutter alike. It was not necessarily those on the list who were 1, 2, 3. Among the top 20, anyone might be a good hire. It's very hard to predict someone's success. Look at me. Maybe I'm a late bloomer, and it could have been that, if I was closed out at an early stage, I wouldn't have had the career I had.  So, I wanted to give those among the top 20 an opportunity.

We hired some very excellent women when I was head of the group, and we attracted people from around the world. We didn't always hire e.g., ``Harvard's best student.'' I thought it was important to give people opportunities. That philosophy of mine goes back to when I was on the admissions committee for graduate students at Stanford.  There were always half a dozen students from ``Harvard'' with straight A's and glowing letters, but many of them never had any experience with hands-on physics except what they did it in a course. The Stanford Physics Department was very much an experimental department, and anybody with experimental experience was valued. It was very important to look for other skills and talents, not just at GPAs. We made an effort not to accept the first straight A's from e.g., Harvard or Caltech but get a mix of people and a mix of interests. Also, I wondered when I was on the admissions committee if I would have admitted myself?   I decided I would have been a borderline case. I probably had very good letters of recommendation from the people I worked with in the experimental labs. But some of my grades were pretty awful, some very good, and some not so good. And in the end I probably wouldn't have admitted myself. That's a historical curiosity. I was on the faculty at Stanford but never accepted as a student there. ``
\end{quote}

In the early 1990's it was becoming clear that the top quark was heavy and that pairs of top quarks may form a composite Higgs boson.   This led Bill, Manfred Lindner, and I to create a theory called   ``top quark condensation'' \cite{BHL},  based upon some 
of my earlier ideas on the renormalization group fixed point of the behavior of the top quark Yukawa coupling \cite{Hill}.   This was one of the first composite Higgs boson theories, and tied together otherwise independent parameters of the standard model.  The theory predicted a heavy top quark, but it also predicted too heavy a Higgs boson.  While the results of the original model proved not to be precisely compatible with the experimental results, the general idea, by incorporating the internal wave-function of the Higgs boson, now agrees well with experiment and makes predictions potentially accessible to the LHC \cite{CTH}.

In 1993, Bill and I recognized that heavy-light mesons, which contain a heavy quark and a light anti-quark, provide a unique window on the chiral symmetry dynamics of a single light quark \cite{Chris1}.  The theory precisely predicted abnormally long-lived heavy-light resonances ten years before their discovery (the $D_s(2317)$ consisting of charm-antistrange at the Babar experiment \cite{Babar}). This was further developed with Estia Eichten \cite{Chris2} and many observables were computed that have been confirmed by experiment. Similar phenomena should be seen in other heavy-light mesons and heavy-heavy-light baryons, predicting a total of 12 long-lived resonances in various systems. 


Bill  felt it was very important that the physics community support the Superconducting Super Collider (SSC) project.  There was some resistance because few people were enthusiastic about going to Texas.   Fred Gilman who was head of SSC research, talked Bill into  becoming head of the new theory group.  He decided it was his obligation to go to Texas to help build a new theory group and help support the project.
Chuck and Karen were now in college, and Marge had a job offer in creating a new education department at the SSC.  
They sold their house in Glen Ellyn, Illinois, and moved to Waxahachee.    

The SSC project was always controversial, there was a lot of debate, and a lot of negative sentiment expressed from various physics communities and others, complaining about the cost of the project. This made it difficult to  encourage people to come.  Some came as visitors, but if you wanted to come as a postdoc it would be an uncertain and nervous time, yet  Bill did manage to hire some postdocs.    But, it was only ten months after the Bardeens had arrived, that the vote by Congress killed the SSC project in October 1993.

Bill's  job was then to help newly hired postdocs to find jobs --- and then find a job for himself!
One of the conditions of going to the SSC from Fermilab was to resign one's Fermilab position since the lab couldn't function with people who went on leave.  When the SSC was killed, that all changed. It took about a month or so to figure out what would actually happen. Marge came back to Fermilab first where she had been the Deputy Head of the SSC Education Office, and the Fermilab Director, John Peoples, wanted her on the Fermilab payroll rather than the SSC payroll.  Bill came back a few months later after all the postdocs got positions and again became acting head for two years (while then Head, Keith Ellis, was away on sabbatical).  They bought a new house in Warrenville, Illinois, and rejoined the Fermilab community.

While  at Fermilab, Bill  had one official sabbatical in 35 years, supported by a Guggenheim Fellowship.  In 1984 -- 1985 he and Marge went, literally, around the world, visiting places, mostly at the invitation of physicists who had been at Fermilab and previously knew him.  Bill always found that visiting them in their space was interesting and became an important part of his career.

 One of the things the Bardeens always did was to hold annual summer picnics for the Theory Group because there were lots of visitors. They had three picnics, one each month, so all visitors could attend one with a potluck barbecue, with games, basketball, and particularly volleyball, and good conversation.  Kids played croquet, shot baskets, etc. There would be as many as 50 people in their backyard. It was good for the families to feel more at home rather than living in an apartment for a month or so.

Bill, of course, believed that our mission is to discover what lies beyond the Standard Model:

\begin{quote} ``There are a number of areas that are important, some of which we work on. How does the Standard Model really work?  Buras, Muta, Duke and I did some of the original work on trying to translate the theory (QCD) into something experiments measure. This has now become much more sophisticated and requires a lot of talent and skill.  Lattice gauge theory is important for the same reason, because some things are not amenable to the usual techniques of perturbative quantum field theory.  Moreover there's a lot of interest in what might be beyond the Standard Model; there's obviously connections with gravity, connections with astrophysics and theory.  There are lot of connections to condensed matter theory as well. In terms of beyond the Standard Model, there are lots of ideas about what nature might be doing. But nature's been clever about hiding it from us up to now. There are certainly big puzzles that the Standard Model doesn't explain both in particle physics and astrophysics.

One of the puzzles for over 100 years has been how do gravity and particle physics fit together, particularly in their quantum nature?  Is it String Theory?  That's all going to be interesting. Some of the new initiatives at Fermilab  are quantum information science and quantum computing. That's a new way of thinking about doing things, and it will be interesting to see how that develops and whether it ends up being useful. Whether you can make a lot of progress on certain problems that you couldn't touch in any other way. It remains to be seen how big an impact they will have in particle physics. A big question is, do we get another big machine to explore beyond what we can see now, or does the field become more a theoretical, conceptual place where we don't get to test things, or we don't have experiments to tell us what to do. Then it becomes a little drier.

 There are a lot of clever ideas in theory that could move the field in a new direction, but we're evidenced based, so at some point, you want to connect with what the experiments see. There are certain discrepancies now. But most of the papers from CMS and ATLAS, for example, are  consistent with the Standard Model. Now, either there are things that don't work, where they're not telling us about, or a lot of the physics that is accessible is in agreement with the Standard Model. In my opinion, there's nothing in the Standard Model that could be the basis to explain all that we see. It doesn't really explain why things are the way they are. There seem to be ``fine tunings'' of parameters where you would say, why did nature do that? Because it requires very subtle relationships. Nevertheless, there's nothing that says that couldn't happen. I seriously would like to explain why it happens and not be satisfied with just saying, well, everything works, but we don't understand why it works. We're able to predict what happens from the theories we have. But we don't really understand why the theory is the way it is.

There's a lot of optimism, a lot of ideas about going forward. It also could be that some of the more formal areas like string theory which had a number of successes every few years, will have another big breakthrough. Our depth of understanding about what theoretical fields are important has slowed down a bit in recent years. But you never know when that could pick up again with new ideas. Or again, as I say, the puzzle of marrying particle physics, gravity, and quantum mechanics is on the minds of many people, it has been for many years. It's a very active area where there could be breakthroughs. But it may take another Einstein.''
\end{quote}

Bill Bardeen had a long career that benefited from interactions around the world and was supported by many groups and fellowships.  He was a ``people person.''  Having the interactions with younger and older people was important to him, especially as someone who stayed active in research as he did for most of his career.  He created and collaborated on a number of very important papers with various colleagues that made a significant difference to theoretical physics.  Here I have cited only the subset of his most cited papers, but
the complete list of works and collaborators is much longer. Bill Bardeen acheived a scientific reputation that transcended
the topics he co-authored, and he became one of the world's leading authorities on real-world
quantum field theory.

In addition to his fellowships, Bardeen was awarded J.J. Sakurai Prize from the American Physical Society in 1996.  He became a Fellow of the American Physical Society in 1984, a Fellow of the American Academy of Arts and Sciences in 1998, a Member of the National Academy of Science in 1999, and Fellow of the American Association for the Advancement of Science in 2009.  Also, reflecting long service, Bardeen was an elected General Member (1990-92) and Trustee (1986-91)  of  the Aspen Center for Physics.  The University of Minnesota awarded him an honorary doctorate in 2002, which if he hadn't finished his thesis might have been his only doctorate!   After suffering a long illness, Bill Bardeen passed away at his home in Warrenville, Illinois, on November 18, 2025.

\section*{Acknowledgments}
I thank Stephen Adler, Marge Bardeen, Chuck Bardeen, and Chris Quigg for many comments and suggestions. 

\end{document}